# Developing a Blockchain-Based Secure Digital Contents Distribution System

Syed Mohiuddin Qadri, Sangwhan Cha
*Harrisburg University of Science and Technology*
Harrisburg, PA, USA
(sqadri, scha)@harrisburgu.edu

*Abstract*— As digital content distribution expands rapidly through online platforms, securing digital media and protecting intellectual property has become increasingly complex. Traditional centralized systems, while widely adopted, suffer from vulnerabilities such as single points of failure and limited traceability of unauthorized access. This paper presents a blockchain-based secure digital content distribution system that integrates Sia, a decentralized storage network, and Skynet, a content delivery network, to enhance content protection and distribution. The proposed system employs a dual-layer architecture—off-chain for user authentication and on-chain for transaction validation using smart contracts and asymmetric encryption. By introducing a license issuance and secret block mechanism, the system ensures content authenticity, privacy, and controlled access. Experimental results demonstrate the feasibility and scalability of the system in securely distributing multimedia files. The proposed platform not only improves content security but also paves the way for future enhancements with decentralized applications and integrated royalty payment mechanisms.

*Keywords—Blockchain, Digital Contents, Contents Distribution, Secure Digital Contents,*

## I. Introduction

Following the commercialization of the internet in 1994 and the rapid evolution of technology, a growing volume of digital content—ranging from text and audio to video, graphics, and animations—has been published and distributed online. Digital content distribution refers to the delivery of such content via digital content delivery platforms. However, challenges related to content security, copyright protection, traceability of copyright violations, and secure distribution have become increasingly significant for both content creators and distributors [1].

According to Qureshi et al. (2021) [1], content protection techniques aim to safeguard digital content from unauthorized access. These techniques encompass mechanisms such as copy protection, traceability, authentication of content sources and recipients, usage control, digital rights management, and secure distribution of both content and access credentials.

Currently, most online digital content protection mechanisms rely heavily on encryption technologies [2]. Despite the adoption of various techniques such as Digital Rights Management (DRM), watermarking, and Conditional Access Systems (CAS), digital content remains vulnerable to a range of decryption attacks, posing ongoing challenges to secure content protection.

Another major challenge lies in the lack of effective traceability within traditional content distribution systems, which are predominantly based on centralized architectures. These centralized systems are prone to single points of failure and make it difficult to trace unauthorized content distribution or downloads.

Blockchain, as a distributed ledger technology, offers a promising alternative by enabling consensus on shared transactional data among untrusted participants in a decentralized network [3]. Its characteristics—such as decentralization, transparency, and immutability—have inspired the development of blockchain-based models for secure content delivery, royalty distribution, C2C transactions, and borderless content consumption.

In this study, we propose and evaluate a blockchain-based system for secure and traceable digital content distribution. Our system leverages the Sia blockchain platform and its Skynet content delivery network. To address limitations inherent in blockchain-based content delivery, we adopt a hybrid architecture consisting of both on-chain and off-chain components. Additionally, we introduce an authentication server to verify the identities of both content owners and consumers within the off-chain layer, enhancing system security and trustworthiness.

Our results demonstrate that the proposed blockchain-based digital content distribution system enables efficient and secure content management and distribution.

The remainder of this paper is organized as follows: Section II provides background information. Section III details the system architecture. Section IV presents and discusses experimental results. Finally, Section V concludes the paper.

## II. Background

Sia is a decentralized cloud storage platform secured by blockchain technology [4]. Originally conceptualized by David Vorick and Luke Champine during HackMIT 2013, the project has since evolved into a fully-fledged development initiative led by the Nebulous team, headquartered across the United States and the European Union. The Sia network leverages unused hard drive space to create a decentralized data storage marketplace that is more cost-effective and reliable than traditional cloud storage solutions.

Sia operates on its own blockchain, powered by a native utility token called Siacoin. To ensure high availability and eliminate single points of failure, data is distributed across multiple geographically diverse locations. This design ensures that data remains private, censorship-resistant, and immune to unauthorized access or manipulation by third parties.

When a file is uploaded to the Sia network, it is automatically divided into segments, encrypted, and distributed to multiple hosts worldwide. The process is fully automated: renters upload files, and hosts provide the storage. The system redundantly replicates files across the network to ensure data availability even in the event of node failures. Importantly, because hosts only receive encrypted file fragments, they are unable to reconstruct or access the original files, thereby ensuring strong data confidentiality.

Previous research has explored the use of blockchain as a security enhancement mechanism in digital rights management (DRM), watermarking, and content protection. For instance, Heo et al. [5] proposed a blockchain-based solution using secret block structures to enhance DRM systems. Yue et al. [6] introduced a blockchain-based architectural framework for multimedia data management.

In contrast to these earlier studies, the present work proposes a comprehensive system architecture for secure digital content distribution using blockchain. Our design not only ensures content protection but also facilitates decentralized, privacy-preserving data distribution leveraging the robustness of blockchain technology.

### III. PROPOSED SYSTEM ARCHITECTURE

As illustrated in Figures 1 and 2, the proposed system architecture consists of two layers: an on-chain layer and an off-chain layer. User authentication is handled in the off-chain layer prior to any transaction to ensure system integrity. While the blockchain allows for public verification of transactions, authentication is a prerequisite to prevent unauthorized or fraudulent activity. All individuals seeking access to digital content must first complete the designated authentication process to ensure secure and trustworthy participation.

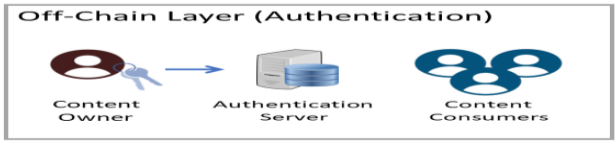

Fig. 1. Off- Chain Layer

Once authenticated, the content provider uploads their content to the server. A consumer who wishes to access the content must also complete the authentication process before submitting a transaction request to the server. Upon receiving the request, the server generates an asymmetric key to enforce Digital Rights Management (DRM) and applies a fingerprinting mechanism using the authenticated user's information to ensure traceability and secure content usage.

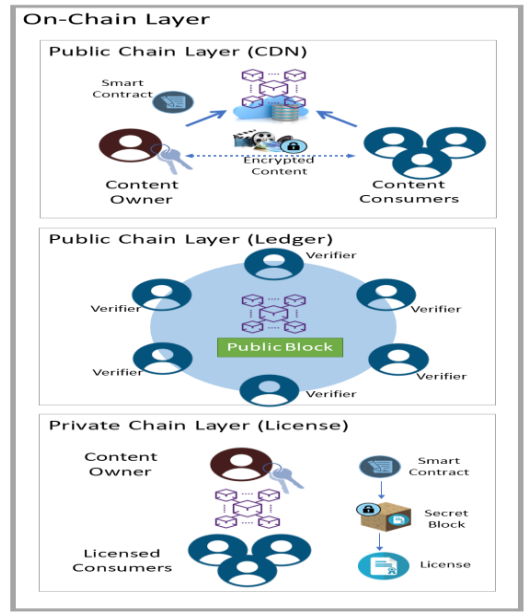

Fig. 2. On-Chain Layer

In the private off-chain component, a license is issued to the consumer for accessing DRM-protected content. This license is accompanied by transaction information, which is encapsulated as a secret block. To receive the license, the consumer initiates a smart contract. The content provider then encrypts the decryption key—used to access the encrypted digital content—using the consumer's public key and transmits it securely via the smart contract. The license, which includes usage policies and access rules, is embedded within the smart contract and delivered to the consumer.

The public on-chain component handles transaction verification and consensus, aggregating validated transactions to construct new blocks. For privacy protection, the secret block containing transaction details is accessible only to the consumer. As a result, transaction verification must be designed to proceed without revealing sensitive transaction information. The structure of the license and secret block is depicted in Figure 3, while the block connection architecture is illustrated in Figure 4.

| License | Hash of the block (hash) | |
|---|---|---|
| Private Key | Previous block hash (public) | Time |
| Key rules : how to use key | User authentication_info | Provider information |
| Information of consumer | Encrypted content information | |
| Rights to use contents | License Information | |
| Hash of the license | | |

Fig. 3. License and Secret Block Structure

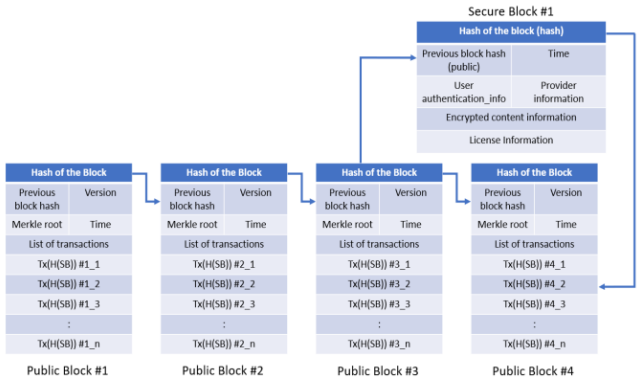

Fig. 4 Block Connection Structure

- User Registration and Authentication

The process begins with user registration. During registration, the user selects an ID and password (PW), generates an asymmetric key pair, and computes a hash value using their password and the hash of their ID. The server stores the user's ID, the resulting password hash, and the public key. When a registered user initiates a transaction, they must first complete the authentication process, as illustrated in Figure 5. Upon receiving an authentication request, the server generates a random challenge value, encrypts it using the user's public key, and sends it back to the user for verification.

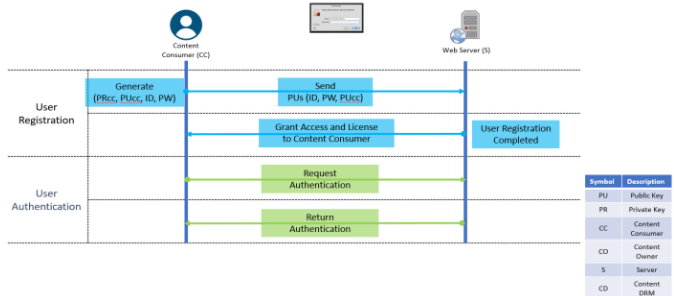

Fig. 5. User Authentication Process

- Encryption on HLS Streaming Protocol

Video streaming involves transmitting large volumes of data to viewers, making the use of uncompressed RAW video files impractical due to their size. To address this, broadcasters encode video content using compression codecs such as H.264 (Advanced Video Coding), significantly reducing file sizes while maintaining quality. One widely adopted streaming protocol is HTTP Live Streaming (HLS), which segments video streams into smaller chunks that can be efficiently transferred over the internet and reassembled by the viewer's video player.

Typically, playback is handled by an HTML5 or Video.js player, which supports native rendering in web browsers. HLS is compatible with a wide range of devices, including smartphones, tablets, laptops, and smart TVs. Because HLS operates over HTTP, it enables broadcasters to use standard web servers or content delivery networks (CDNs) to store and deliver video content at scale.

A key component of HLS is the manifest file, commonly referred to as the M3U8 playlist. This file guides the video player in selecting and retrieving the appropriate video segments for adaptive bitrate (ABR) streaming. Additionally, the M3U8 manifest may include references to the encryption keys used to secure each video segment, ensuring both efficient delivery and content protection.

IV. EXPERIMENTAL RESULTS

We conducted a series of experiments involving the upload and download of multimedia files using our proposed system, following the steps outlined below.

1. Installing SkyNet CLI

```
$ go install github.com/SkyNetLabs/skynet-cli/v2/cmd/skynet@latest
```

```
qadri@SA02-1234567:~/go$ go install github.com/SkyNetLabs/skynet-cli/v2/cmd/skynet@latest
go: downloading github.com/SkyNetLabs/skynet-cli/v2 v2.1.1
go: downloading github.com/SkyNetLabs/skynet-cli v1.1.0
go: downloading github.com/SkyNetLabs/go-skynet/v2 v2.1.0
go: downloading gopkg.in/yaml.v2 v2.3.0
```

2. Encrypt the Content using Google's Shaka Packager and Widevine DRM Solution

```
$ packager in=.\dolby-vision-art.mp4,stream=audio,output=.\dolby-vision-art-audio.mp4
    in=.\dolby-vision-art.mp4,stream=video,output=.\dolby-vision-art-video.mp4
    --enable_widevine_encryption
    --key_server_url https://license.uat.widevine.com/cenc/getcontentkey/widevine_test
    --content_id 7465737420636f6e74656e74206964
    --signer widevine_test
    --aes_signing_key 1ae8ccd0e7985cc0b6203a55855a1034afc252980e970ca90e5202689f947ab9
    --aes_signing_iv d58ce954203b7c9a9a9d467f59839249
    --protection_systems Widevine,PlayReady
    --mpd_output .\dolby-vision-art-drm.mpd
```

3. Upload File to SkyNet

```
$ ./skynet upload ~/Videos/file_example_MP4_1920_18MG.mp4
```

```
qadri@SA02-1234567:~/go/bin$ ./skynet upload ~/Videos/file_example_MP4_1920_18MG.mp4
Successfully uploaded file! Skylink: sia://AADjq68QKeH_I5Aha6hQgF8nWawaQfV_HhYbZhkEwaHk3g
```

When a file is uploaded to Skynet, its contents are hashed using a cryptographic function. This hash, along with supplementary metadata, is used to generate a unique identifier known as a *skylink*. Because the skylink is derived from the file's hash, any modification to the file results in a different hash and, consequently, a different skylink. This mechanism ensures data integrity: Skynet can verify that the file retrieved from storage hosts has not been tampered with, as even a minor change would produce a mismatched hash. For this reason, files on Skynet are immutable—any attempt to alter a file's contents necessitates the creation of a new skylink.

4. Download File from SkyNet

```
$ ./skynet download sia://AADjq68QKeH_I5Aha6hQgF8nWawaQfV_HhYbZhkEwaHk3g ~/Videos/myVideoDownload.mp4
```

```
qadri@SA02-1234567:~/go/bin$ ./skynet download sia://AADjq68QKeH_I5Aha6hQgF8nWawaQfV_HhYbZhkEwaHk3g ~/Videos/myVideoFrom
Sia.mp4
Successfully downloaded skylink!
```

5. Accessing the Content Securely from Internet Browser

```
https://siasky.net/AADjq68QKeH_I5Aha6hQgF8nWawaQfV_HhYbZhkEwaHk3g
```

## V. CONCLUSIONS AND FUTURE WORK

This paper presented a platform architecture for a blockchain-based secure digital content distribution system, aimed at addressing issues of single points of failure and copyright infringement inherent in centralized systems. We began by identifying the core services essential to the architecture, and subsequently introduced a two-layer design consisting of an off-chain data layer and an on-chain data layer.

To ensure secure access, we proposed an access control mechanism that incorporates the concept of self-sovereign identity for decentralized identity management. The feasibility of the proposed solution was demonstrated through a practical use case, and quantitative experiments were conducted to evaluate its performance. The results indicate that the system is not only functional but also scalable.

As future work, we plan to implement the on-chain data layer using a Solana-based decentralized application (dApp) to enhance the overall decentralization of the system. Additionally, we aim to extend the architecture with an integrated payment and royalty distribution system.

## REFERENCES

[1] A. Qureshi and D. M. Jimenez, "Blockchain-Based Multimedia Content Protection: Review and Open Challenges," Appl. Sci. 2021, 01-Nov-2021. [Online]. Available: https:// dx.doi.org/10.3390/app11010001. [accessed: 05-July-2022].

[2] J. Kishigami, S. Fujimura, H. Watanabe, A. Nakadaira and A. Akutsu, "The Blockchain-Based Digital Content Distribution System," 2015 IEEE Fifth International Conference on Big Data and Cloud Computing, 2015, pp. 187-190

[3] Y. Liu, Q. Lu, C. Zhu, et al., "A Blockchain-based platform architecture for multimedia data management," Multimedia Tools and Applications, vol. 30, pp. 707–723, 2021

[4] Sia, Decentralized data storage [Online] available : https://sia.tech/, [Accessed: 25-August-2022].

[5] G. Heo, D. Yang, I. Doh and K. Chae, "Efficient and Secure Blockchain System for Digital Content Trading," in IEEE Access, vol. 9, pp. 77438-77450, 2021, doi: 10.1109/ACCESS.2021.3082215.

[6] Liu, Yue & Lu, Qinghua & Zhu, Chunsheng & Yu, Qiuyu. (2020). A Blockchain-based Platform Architecture for Multimedia Data Management.